%% file: paper.tex
\documentclass[fleqn,usenatbib]{mnras}

\usepackage{newtxtext,newtxmath}
\usepackage{multirow}
\usepackage{multicol}
\usepackage[normalem]{ulem}
\usepackage[T1]{fontenc}
\usepackage{ae,aecompl}

\usepackage{graphicx}	

\newcommand{\numax}{\mbox{$\nu_{\rm max}$}}
\newcommand{\Teff}{\mbox{$T_{\rm eff}$}}

\newcommand{\mosc}{\mbox{$\langle P_{\rm osc}\rangle$}}
\newcommand{\mgran}{\mbox{$\langle P_{\rm gran}\rangle$}}
\newcommand{\bnumax}{\mbox{$B_{\numax{}}$}}

\newcommand{\hosc}{\mbox{$H_{\rm osc}$}}

\newcommand{\kepler}{{\em Kepler\/}}

\DeclareMathOperator{\sinc}{sinc}

\newcommand{\Dnu}{\mbox{$\Delta\nu$}}

\newcommand{\new}[1]{{\color{red}{#1}}}
\newcommand{\delete}[1]{\sout{#1}}
\renewcommand{\new}[1]{#1}
\renewcommand{\delete}[1]{}

\usepackage{longtable}
\usepackage{booktabs} 
\graphicspath{{./}{figures/}}
\newif\ifarxiv
\arxivfalse 

\title[Oscillations and granulation with Kepler and \em TESS]{Testing the wavelength dependence of oscillations and granulation in red giants using \kepler{} and TESS}

\author[Sreenivas et al.]{%
K. R. Sreenivas$^1$\thanks{E-mail: skal9597@uni.sydney.edu.au} ,Timothy R. Bedding$^1$,
Daniel Huber$^{1,2}$,
Courtney L. Crawford$^1$,
\newauthor
Dennis Stello$^3$,
May G. Pedersen$^1$,
Yaguang Li$^{2}$,
and 
Daniel Hey$^{2}$
\\
$^1$Sydney Institute for Astronomy, School of Physics, University of Sydney, NSW 2006, Australia. \\
$^2$ Institute for Astronomy, University of Hawai`i, 2680 Woodlawn Drive, Honolulu, HI 96822, USA\\
$^3$School of Physics, University of New South Wales, Sydney, NSW 2052, Australia.\\
}

\date{}

\pubyear{2024}

\begin{document}
\label{firstpage}
\pagerange{\pageref{firstpage}--\pageref{lastpage}}
\maketitle

\begin{abstract}
Stellar oscillations and granulation in red giants are both powered by convection. Studying the wavelength dependence of their amplitudes can provide useful insights on the driving mechanism. It is also important for plans to carry out asteroseismology with the Nancy Grace Roman Space Telescope, which will operate in the near infrared, to check the dependence of oscillations and granulation on the observational wavelength.
In this work, we aim to understand how the oscillation and granulation power in red giants depend on the wavelength and study how existing predictions compare with observations. 
We measure the mean oscillation and granulation power of 279 \kepler{} red giants, from the power density spectra derived using \kepler{} PDCSAP and TESS-SPOC light curves. 
We find that selection of light curves is important for the study of amplitudes, since different light curve products from TESS show different values of amplitudes. We show that the oscillation and granulation power ratios between TESS and \kepler{} match the theoretical prediction, confirming that both decrease as we move to redder wavelengths. We also see that the mean ratios of oscillations and granulation agree, suggesting that oscillation and granulation have the same wavelength dependence. We also find that the mean height-to-background ratio for \kepler{} agrees with previous results and shows good agreement with TESS. These results suggest that the granulation signals would not severely affect the detection of oscillations. We checked the dependence of these ratio between \kepler{} and TESS on stellar parameters, and see no trends.
\end{abstract}





\section{Introduction}

Oscillations and granulation are two extensively studied phenomena in red giants. Oscillations are thought to be stochastically excited and damped by the convection in the outer layers \citep{goldreich1977a,goldrieich1977b,hekkerjcd, houdek+dupret2018, basuhekkerreview, zhou2020}. Analysis of the oscillation mode frequencies, amplitudes and linewidths allows us to probe stellar interiors and retrieve global properties of stars \citep{chaplin2013, jack2021}. Stellar granulation, an intensity fluctuation predominant at low frequencies, is also a result of convection. The hot plasma from the outer convective zone rises to photosphere resulting in irregularly shaped bright cells (granules), which cool to produce darker regions. This granulation produces brightness fluctuations whose power and time scale depends on the surface properties of the star \citep{tpdh2001, ludwig2009,mathur2011, kjb11, kallinger2014, enrico2017,rodrigran}. 

 Large volume of observational data from CoRoT \citep{augergne2009, baglincorot2006} and \kepler{} \citep{koch2010, Borucki2010} were used to check the validity of \new{energy transport theories and  scaling relations of solar like oscillators proposed by various authors \citep{goldreich1977a, goldrieich1977b, jcd1983, murray1993, kjb95}}. While calculating the theoretical oscillation amplitudes, most of these studies assumed an adiabatic nature for stellar oscillations and found discrepancies between the \new{observed and} predicted amplitudes calculated using scaling relations \citep{Huber_2011, stello12, vrard2018}. Although \cite{samadi2012} accounted for the non-adiabatic effects, their study could not fully reproduce the observed amplitudes, only reduce the disagreement to 40\,\%. On the other hand, various studies have investigated the dependence of \new{observed} granulation power on stellar parameters and are in mutual agreement \citep{mathur2011,kallinger2014, enrico2017, rodrigran}. All of these studies focused on the relationship between stellar parameters and amplitudes of solar-like oscillators. 
\begin{figure}
 \includegraphics[width = \linewidth]{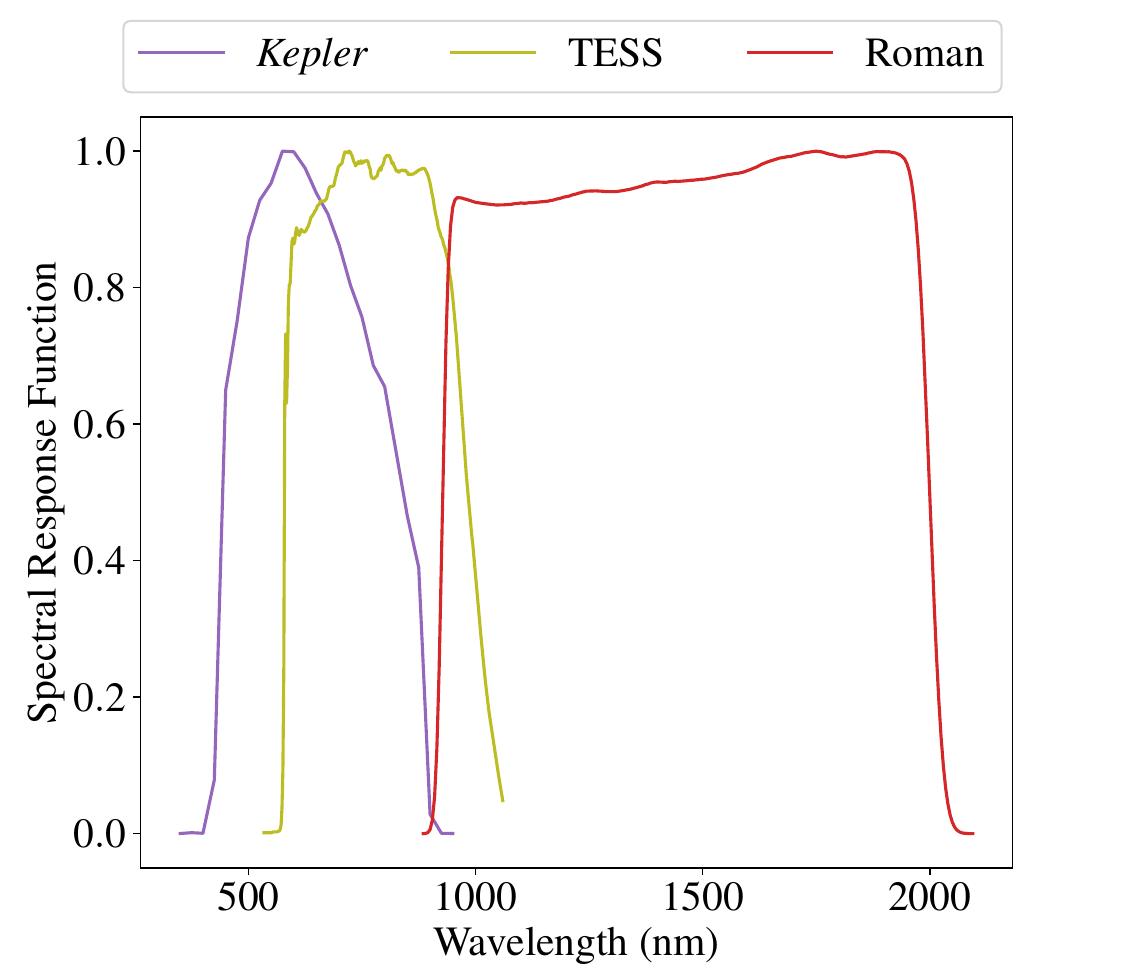}
 \caption{Spectral response function of \kepler{}\,\citep{keplerresponse}, TESS\,\citep{ricker2014} and the wide-field instrument onboard the Nancy Grace Roman Space Telescope\,\citep{romanwfirst}. }
 \label{fig:filters}
\end{figure}

By comparing \kepler{} data with those from TESS (Transiting Exoplanet Survey Satellite; \citealt{ricker2014}), which have different but overlapping pass bands,  it is possible to test predictions of how oscillation amplitudes and granulation power vary with wavelength. 
With upcoming launch of \new{the} Nancy Grace Roman Space Telescope, which observes in the near-infrared, we will be probing oscillations in red giants in a completely new wavelength regime \citep{gould2015,huber2020}. 
Figure\,\ref{fig:filters} shows the response curves of \kepler{}, TESS and Roman. 

\cite{kjb95} noted that the amplitude of oscillations decreases towards longer wavelengths. A theoretical study by \cite{lund2019} showed that the amplitude of oscillation across the two pass bands varies with the ratio of their bolometric correction factor. \cite{lund2019} calculated bolometric correction factors through two separate methods: one by assuming a Planck spectrum and the other using synthetic stellar spectra. The ratio of oscillation amplitudes between \kepler{} and TESS was predicted to be around 0.83 (0.69 in power). Further, it has been observed that the region of oscillations in the frequency domain contain some level of granulation signal, hence is often considered as noise \citep{huber2009, mosser2012, mathur2011}. Therefore, a proper analysis regarding the behaviour of oscillation and granulation among different stellar types is relevant to understand the wavelength sensitivity and thus \new{formulate strategies to account for the effect of granulation component}.  

In the Sun, \new{various studies \citep{michel2009, ballot2011, salabert2011, García_2013, Lund_2014, sulis2020a} have shown that} the oscillation and granulation levels decrease as a function of wavelength, using VIRGO time series \citep{Fröhlich1995}. 
%
In this paper, we compare the observed oscillation and granulation amplitude ratios with predictions for 279 \kepler{} red giants that were observed by both \kepler{} and TESS. 


\section{Data, target selection and methodology}

 \subsection{Light curves}


From the \kepler{} mission, we used the Long Cadence (29.4 min) light curves  spanning the full mission. The light curves were processed through the Pre-search Data Conditioning Simple Aperture Photometry (PDCAP) algorithm at the \kepler{} Science Operations Center (SPOC; \citealt{pdcsap12012,pdcsap22012}). From TESS, we used light curves derived from Full-Frame Images (FFIs) using the SPOC algorithm (TESS-SPOC; \citealt{spoc2016, tessspoc}). We did not apply any additional \new{high-pass} filtering to the light curves, \new{as we found that it causes a change in amplitudes by 0.2$\pm$2.0\,\% which could be significant in our analysis.}

SPOC is one of several pipelines that have been used to create light curves from TESS FFIs. These pipelines select the aperture used for extracting the light curve in different ways. Also, correction for systematic effects, such as background light, is carried out differently, which could affect the measurement of the amplitudes. Therefore, in order to check the amplitudes measured by TESS-SPOC, we examined a sample of contact binaries, as detailed below.


\subsection{TESS pipeline comparisons using contact binaries}
\label{cbs}
\new{Checking amplitude measurements would ideally be done with an astrophysical object that varies by the same amount in the \kepler{} and TESS pass bands.  However, all astrophysical phenomena have some wavelength dependence. We chose to use contact binaries, which show near sinusoidal variations in their light curve due to tidal distortion. Their amplitudes only differ by small amounts as a function of wavelength, depending on the effective temperature of the components \citep{Hedges_2021}. We discuss this effect below, at the end of the section.} 

We selected 51 \kepler{} contact binaries from \cite{prsa11} that have TESS-SPOC light curves. The periods of these systems are in the range 0.1 to 2.6\, d. To check whether the extraction of light from FFIs and de-trending method influences the amplitudes, we also examined the TESS light curves processed using QLP pipeline (Quick Look Pipeline; \citealt{qlp1, qlp2}) and the TGLC pipeline (TESS--Gaia Light Curve), which uses the Gaia catalogue as a prior to derive the light curve \citep{tglc}. The TGLC pipeline produces two different light curves, one by modelling the point-spread function (PSF) of FFIs and one using aperture photometry (APER). Since the TGLC data products for \kepler{} stars are only available for sectors 14 and 15, we only used these two sectors of data from all the above light curve products. In addition, we ensured that all time series have the same time span of observations as TESS-SPOC. 
\begin{figure}
 \includegraphics[width = 1\linewidth]{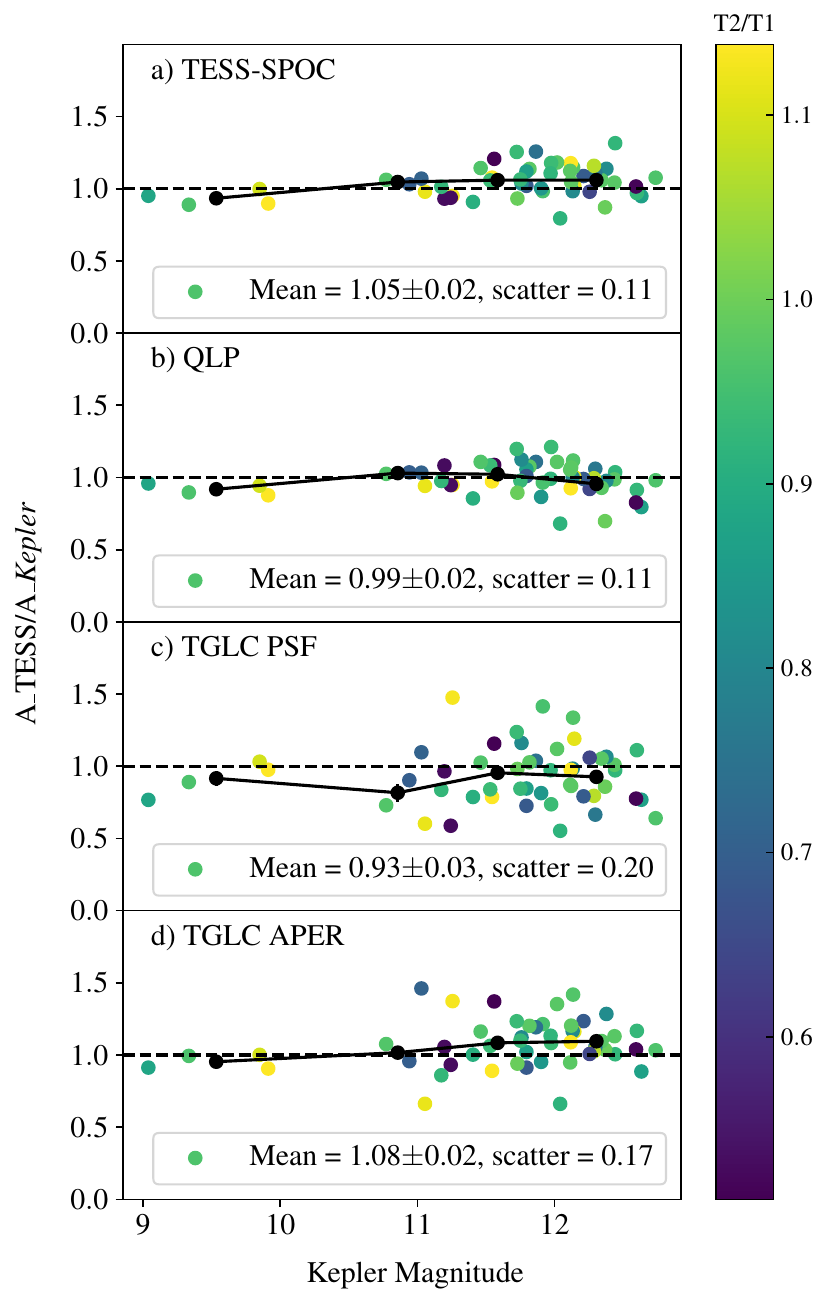}
 \caption{Ratio of amplitudes of 51 contact binaries in TESS relative to \kepler{} for various sources of light curves, as a function of \kepler{} magnitude. Panel a with TESS-SPOC light curves, panel b for QLP, panel c for TGLC PSF and panel d for TGLC APER light curves. The points are colour-coded with the temperature ratio between the secondary and primary components. Error bars show the standard error on the mean (SEM) in each bin and each legend gives the overall mean, SEM and standard deviation.}
 \label{fig:binary}
\end{figure}
 For each star, we calculated the amplitude spectrum of the light curve from each of four the pipelines. We measured the amplitude of the highest peak from \kepler{} and then, at the same frequency in each of the TESS light curves. It should be noted that the highest peak was at half of the reported period of binary orbit because the two components have very similar tidal distortions. 

 Figure \ref{fig:binary} shows the ratio of amplitudes of TESS to \kepler{} at the measured frequency, as a function of \kepler{} magnitude. The individual points are colour-coded with the temperature ratio between the components, taken from \cite{prsa11}. We do not see any systematic trend in the ratios as a function of temperature ratios \new{or the mean effective temperatures of the binary.}
 
 \new{In order to understand the magnitude of variation caused by the different temperatures of the components, we used PHOEBE \citep{phoebep1} to generate synthetic \kepler{} and TESS light curves of contact binaries for a grid of temperature ratios and mass ratios covering a range similar to the observed systems. We measured the amplitudes from these synthetic light curves in the same was as for the observations. We find that the ratio between the amplitude of binaries in TESS and \kepler{} varies by 0.4 to 1.2\,\%, depending on the temperature ratios and mass ratios. This spread is smaller than the scatter on amplitudes from quarter-to-quarter variations of \kepler{}  data alone, which we measured to have a median value of 1.4\,\%. In addition to this, the amplitudes measured from different TESS light curves have a pipeline-to-pipeline scatter ranging from 5\,\% to 21\,\% for the stars in our sample.} This shows that the \new{measured} scatter in amplitude ratios shown in Fig. \ref{fig:binary} is not due to the temperature difference between components, but is instead dominated by systematics from different pipelines and instrumental effects. 
 
 It can be seen \new{from Fig. \ref{fig:binary}} that the TESS pipelines show reasonable agreement with amplitudes from \kepler{}. The amplitude ratios from the SPOC-generated light curves have similar scatter to those from QLP light curves and the scatter from TGLC is greater. This test confirms that TESS-SPOC light curves are suitable for this project. Moreover, the measured scatter of 0.11 on the ratios can be viewed as the minimum scatter to be expected when comparing the oscillation and granulation ratios between \kepler{} and TESS-SPOC, in addition to the variability due to the stochastic nature of oscillations and granulation. 


\subsection{Sample selection of red giants}
 \begin{figure}
 \includegraphics[width = \linewidth]{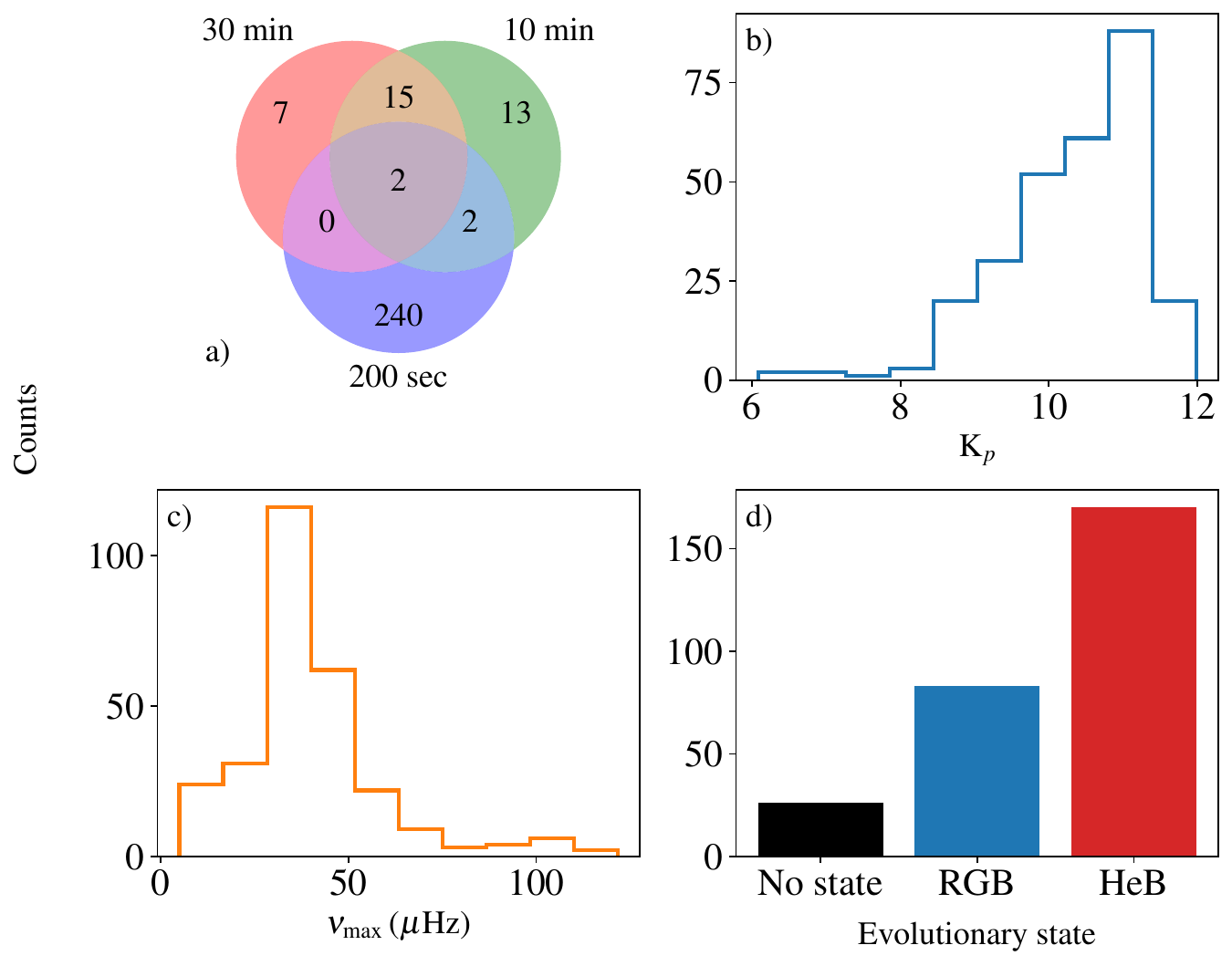}
 \caption{Distribution of various parameters for the 279 stars in our sample. (a)~The cadence distribution of TESS-SPOC data. (b)~The distribution of \kepler{} magnitudes. (c)~The distribution of \numax{} from \citet{sreenivas2024}. (d)~The distribution of evolutionary states from \citet{yu2018}. }
 \label{fig:dist}
\end{figure}
The asteroseismic characteristics of \kepler{} red giants have been thoroughly examined in several catalogues \citep{Huber_2011, mosser2012, kallinger2014, vrard2018, yu2018}. These catalogues contain estimates of the frequency of maximum power (\numax{}; \citealt{brown1991,kjb95}) and the large frequency separation (\Dnu; \citealt{tassoul1980asymptotic}), determined by analysing frequency spectra using dedicated pipelines \citep{hekker2012, mathur2011}.

\cite{stello22} carried out a magnitude-limited analysis (Kepler magnitude less than 13) of \kepler{} red giants from \cite{yu2018}, using two sectors of TESS light curves. This resulted in a subset of stars which showed a clear power excess with TESS at the previously reported \kepler{} \numax{}. They measured \numax{} values for 2724 stars and \Dnu{} values for 570 stars from the TESS data, using the SYD pipeline \citep{huber2009}. For this study, we used the 226 Kepler red giants from \cite{stello22} that have at least one sector of TESS-SPOC data (we omitted two stars that have \kepler{} magnitudes fainter than 12). We also included 53 \kepler{} red giants from \cite{yu2018} that were not studied by \cite{stello22} but which show clear oscillations in TESS-SPOC.

Figure \ref{fig:dist} shows the distribution of parameters for the 279 stars in our sample. From panel a, most of the stars (about 90\,\%) are from the second extended mission of TESS, with FFI light curves having a cadence of 200\,s. Note that, 260 stars in this sample have only one sector of TESS data. Panel b shows the distribution of the \kepler{} magnitude of the stars. From Fig. \ref{fig:dist}\,c, it can be seen that most stars in our sample have \numax{} values $\sim$ 30 to 70 $\mu$Hz and none are above 124\,$\mu$Hz. This is because of two reasons: the non-availability of TESS-SPOC light curves for a larger sample and the inability of TESS, with its much smaller aperture, to detect oscillations at higher \numax{} with short datasets \citep{Hon_2021,stello22}. Fig\,\ref{fig:dist}\,d shows the evolutionary state of the red giants in our sample. About 70\,\% of the stars burn helium in their core (HeB).

\subsection{Methodology}

Since our aim is to compare oscillations and granulation of the same star observed with \kepler{} and TESS, which have different time spans and cadences, it is important that we do our analysis in power density. Analysis in power density units ensure that the measured quantities are independent of the time span of observations \citep{kjb95, kjb2005}. For \kepler{}, we calculated the power density spectra of the full light curves up to the Nyquist frequency (283 $\mu$Hz). For TESS data, because the observational cadence was sometimes different between sectors, we first constructed the power spectra (in ppm$^{2}$) of each sector \citep{lomb1976,scargle1982, press1989}, up to the lowest Nyquist frequency ($\nu _{\rm Nyq}$) among all sectors. These were converted to power density (in ppm$^{2}$/$\mu$Hz) by multiplying each by the effective observation time, which is the number of points multiplied by the cadence of that sector. 

\begin{figure}
 \includegraphics[width = \linewidth]{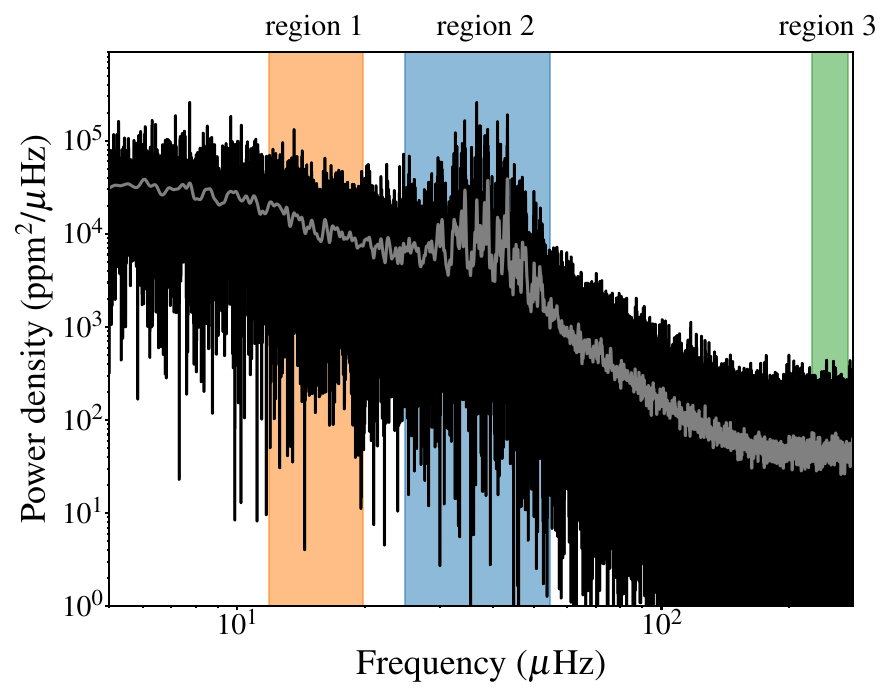}
 \caption{Power density spectrum of the \kepler{} light curve for a typical red giant star, KIC\,10341704. The grey curve is the power spectrum after smoothing by 0.05\,$\rm \Delta\nu$ (0.22\,$\mu$Hz). The three regions are discussed in Sec.~\ref{subsectmethod1}.}
 \label{fig:ampmeas}
\end{figure}

The averaging of signal during each integration causes the attenuation of amplitudes \citep{huber2009, Chaplin_2011,kallinger2014}. Correction for this is important since we are comparing amplitudes from \kepler{} and TESS, where the exposure times are usually different. To correct for this, we first calculated the mean high-frequency noise from $0.8\nu _{\rm Nyq}$ to $0.97 \nu _{\rm Nyq}$ (see region 3 in Fig. \ref{fig:ampmeas}). We subtracted this high frequency noise from the power density spectra and divided by $\sinc^{2}\left(\frac{\pi}{2}\frac{\nu}{\nu _{\rm Nyq}} \right)$. We then averaged the power density spectra from individual sectors, resulting in the final power density spectra of the star. 

Next, we describe two different methods to calculate the granulation and oscillation power, as detailed below.

\subsubsection{Method 1: Using Mean powers}
\label{subsectmethod1}
In order to measure the oscillation and granulation power, we first employed a simple and uniform approach. We used the \numax{} values from \cite{sreenivas2024} to define the regions where the oscillation and granulation are prominent. We defined 3 regions of interest in the power density spectrum of each star: region 1 for granulation, region 2 for oscillations and region 3 for measuring white noise. Figure \ref{fig:ampmeas} shows an illustration of these regions for a typical \kepler{} star.

As mentioned above, the white noise was measured in region~3, from $0.8\,\nu_{\rm Nyq}$ to $0.97\,\nu _{\rm Nyq}$. Previous studies have shown that the power becomes purely granulation in nature between 800\,$\mu$Hz to 1000\,$\mu$Hz for the Sun (\citealt{Karoff_2013} and references therein), which spans from 0.26 to 0.32 times \numax{}$_{\odot}$. To get accurate measurements from the shorter time span of TESS data, we defined a wider granulation region than this, from 0.3\,\numax{} to 0.5\,\numax{}, and the mean power \mgran{} in this region 1 was used for granulation. This choice ensures that the measured power is free of long term stellar signals at lower frequencies and is purely due to stellar granulation. 

We measured the mean oscillation power, \mosc{}, in region~2, with boundaries from $\numax - \delta\nu_{\rm env}$ to $\numax + \delta\nu_{\rm env}$. We set $\delta\nu_{\rm env}$ to equal the full-width at half-maximum (FWHM) of the power excess envelope, which can be estimated from its power law relation with \numax{} \citep{mosser2012, stello22}. To find the coefficients in this power law, we fitted the FWHM values measured by \citet{yu2018}, using their uncertainties as weights in the fit and only including stars for which the FWHM was measured to better than 10\%. The resulting fit gave the FWHM as $0.78 (\numax/\mu{\rm Hz})^{0.80}$. It should be noted that region 2 also includes a contribution from granulation, but still gives a useful measure of oscillation power. 

\begin{figure}
 \includegraphics[width = \linewidth]{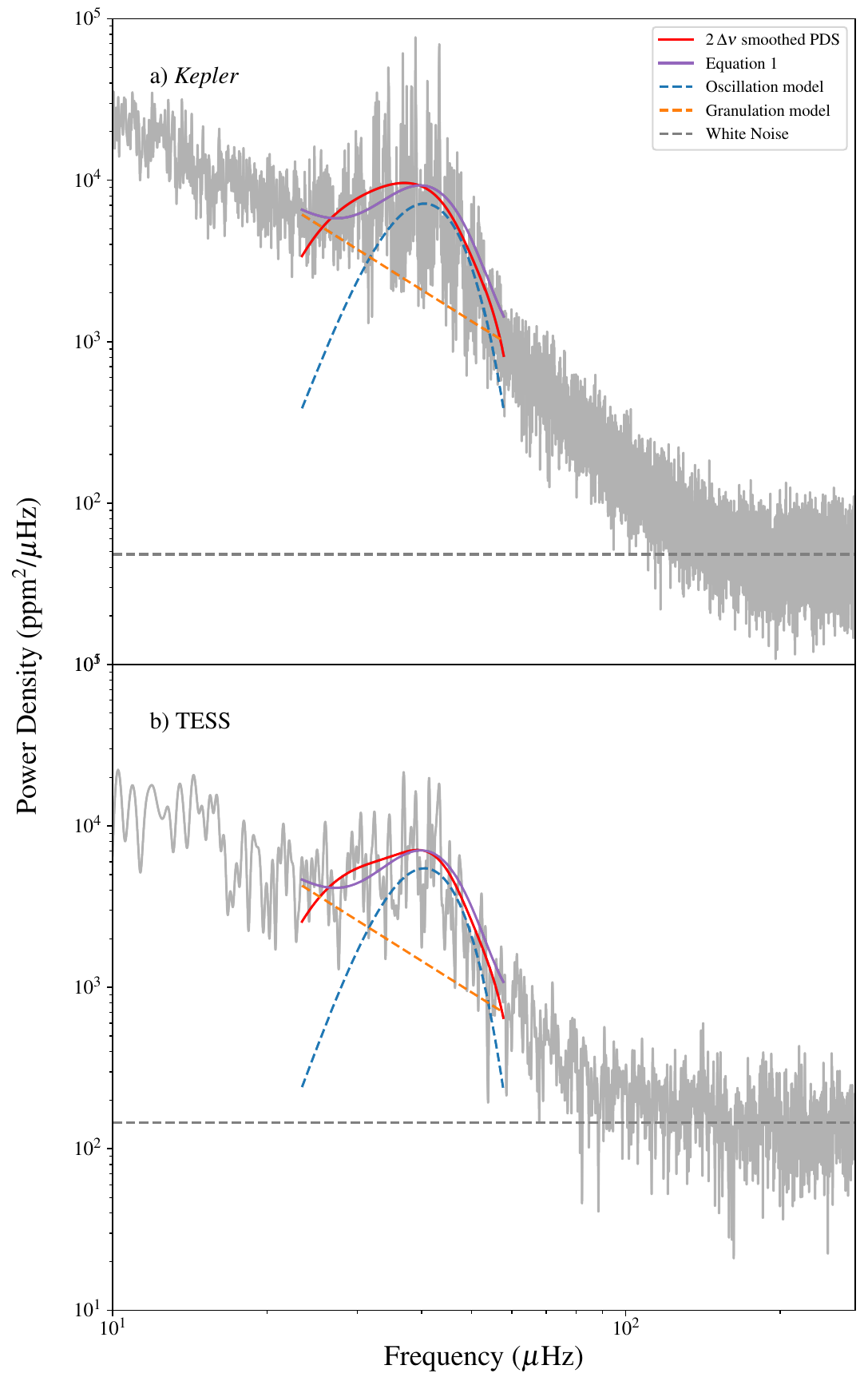}
 \caption{Fit to the power density spectrum of KIC\,10341704 to determine the oscillation and granulation amplitudes, as described in Sec.~\ref{subsectmethod2}. The red line shows the smoothed power density spectrum on which eq. \ref{eq:eq1} is fitted (purple line). The blue and orange dashed lines shows the oscillation component and background component, respectively. Dashed grey lines represent the white noise. a)~ for \kepler{} b)~for TESS.    }
 \label{fig:fit}
\end{figure}

\begin{figure}
 \includegraphics[width = \linewidth]{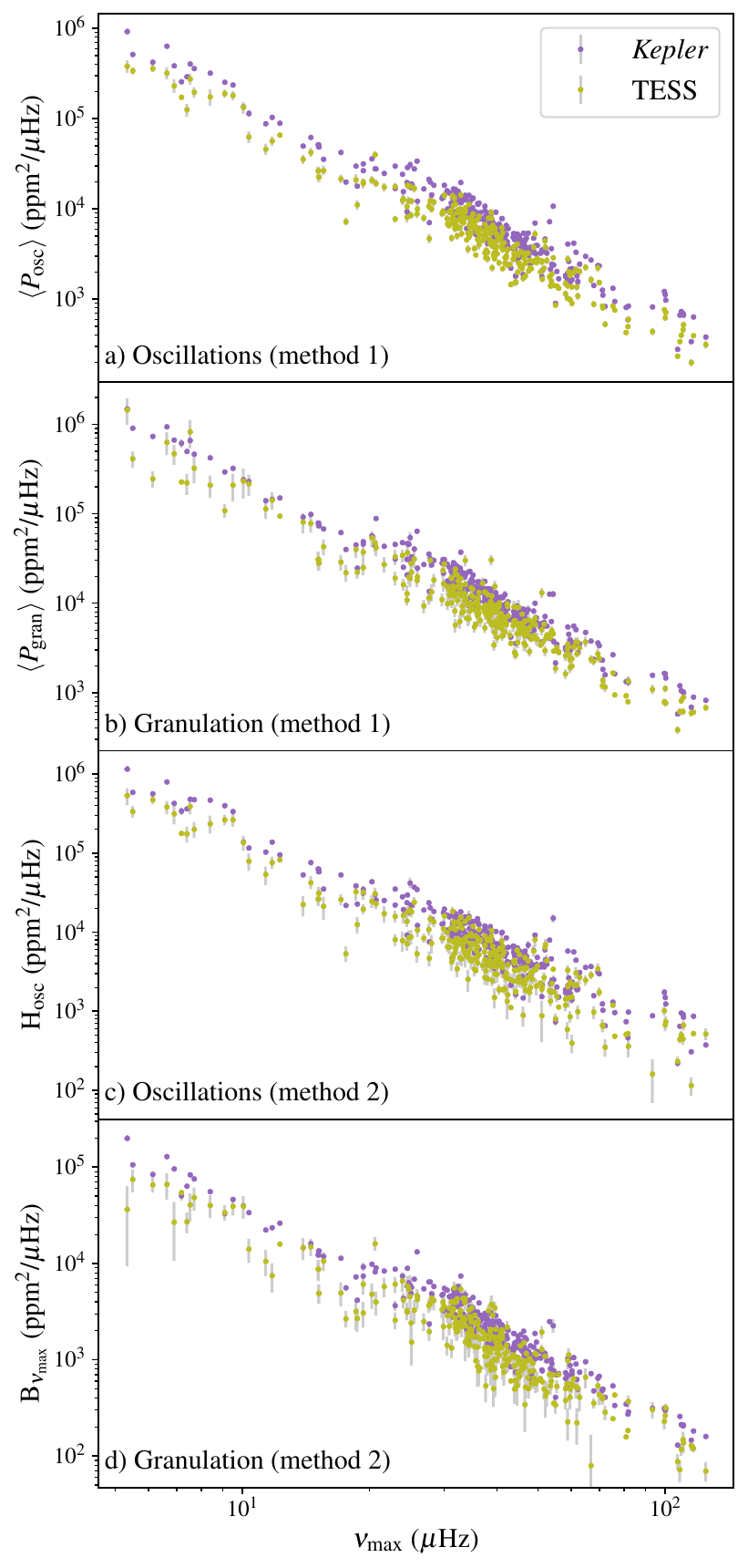}
 \caption{ Oscillation and granulation power determined using the two methods, for \kepler{} (purple) and TESS (yellow). a)~Mean oscillation power,\,\mosc{}.  b)~Mean granulation power,\,\mgran{}. c)~Oscillation power at \numax{},\,\hosc{}. d)~Granulation power at \numax{},\,\bnumax{}. \new{Error bars are shown in grey.}}
 \label{fig:powers}
\end{figure}

\subsubsection{Method 2: Modelling the power excess}
\label{subsectmethod2}
Although the method described in Sec. \ref{subsectmethod1} is straightforward, it does not account for granulation power in the region of oscillations. Since we are focused on the region of power excess, 
we also determined the granulation power at \numax{} (\bnumax{}) and oscillation power (\hosc) using a fitting procedure in the region of oscillations \citep{mosser2012, gehanfra}. \new{In order to fit the power density spectra using Gaussian statistics \citep{huber2009},} we first smoothed the power density spectrum using a Gaussian kernel with a width of $2\Dnu_{\rm exp}$, where $\Dnu_{\rm exp}$ is an estimate of the expected large frequency separation using the relation $\Dnu_{\rm exp}$ = $0.26\,\numax{}^{0.778}$ \citep{stello2009}. In order to isolate the power excess, we fitted to region 2 (Sec. \ref{subsectmethod1}). We fitted the following equation to this smoothed spectrum:
\begin{equation}
 P({\nu}) = \bnumax \Bigg( \frac{\nu}{\numax{}} \Bigg)^{\alpha} + \hosc \exp\Bigg({\frac{-(\nu - \numax{})^{2}}{2\sigma^{2}}}\Bigg) + W \,\,\,.
\label{eq:eq1}
\end{equation}\\
Here, $\alpha$ is the exponent that controls the shape of the background and we choose to fix $\alpha = -2$ (following \citealt{mathur2011, mosser2012, kallinger2014, sreenivas2024}); $\sigma$ is the standard deviation of the oscillation envelope, which is $\delta\nu_{\rm env}$/2.35; \hosc{} is the power density  at \numax{}; and $W$ is the high frequency noise in the power density spectra, which we measured as per section \ref{subsectmethod1} and kept constant at this value throughout the fitting process. Figure \ref{fig:fit} shows the fit to the power density spectrum of a red giant, using \kepler{} data (panel a) and TESS data (panel b). 

\begin{table}
 \centering
 \begin{tabular}{|c|c|c|c|}
 \hline
  Mission & parameter & $\new{\alpha}$ & $\new{\beta}$\\ 
  \\
   \hline
  \multirow{4}{4em}{\kepler{}} &\mosc{} & 0.58$\pm$0.06 & $-$2.37$\pm$0.04 \\ 
  &\mgran{}{} & 0.81$\pm$0.05 &$-$2.37$\pm$0.03 \\
  & \hosc{} & 0.70$\pm$0.07 & $-$2.42$\pm$0.04\\
  &\bnumax{} & 0.80$\pm$0.05 & $-$2.19$\pm$0.03 \\
  \hline
  
  \multirow{4}{4em}{TESS} & \mosc{} &0.32$\pm$0.06 & $-$2.32$\pm$0.06 \\ 
  &\mgran{}{} &0.51$\pm$0.06 & $-$2.28$\pm$0.04 \\
  & \hosc{} & 0.42$\pm$0.09 & $-$2.37$\pm$0.05 \\
  &\bnumax{} &0.53$\pm$0.08 & $-$2.14$\pm$0.05 \\
  \hline
 \end{tabular}
 \caption{Table of fitted parameters, when a model of the form $\alpha \numax{}^{\beta}$ was fitted to oscillation and granulation powers.}
 \label{tab:tab1}
\end{table}

\section{Results and Analysis}

\begin{figure*}
 \includegraphics[width = \linewidth]{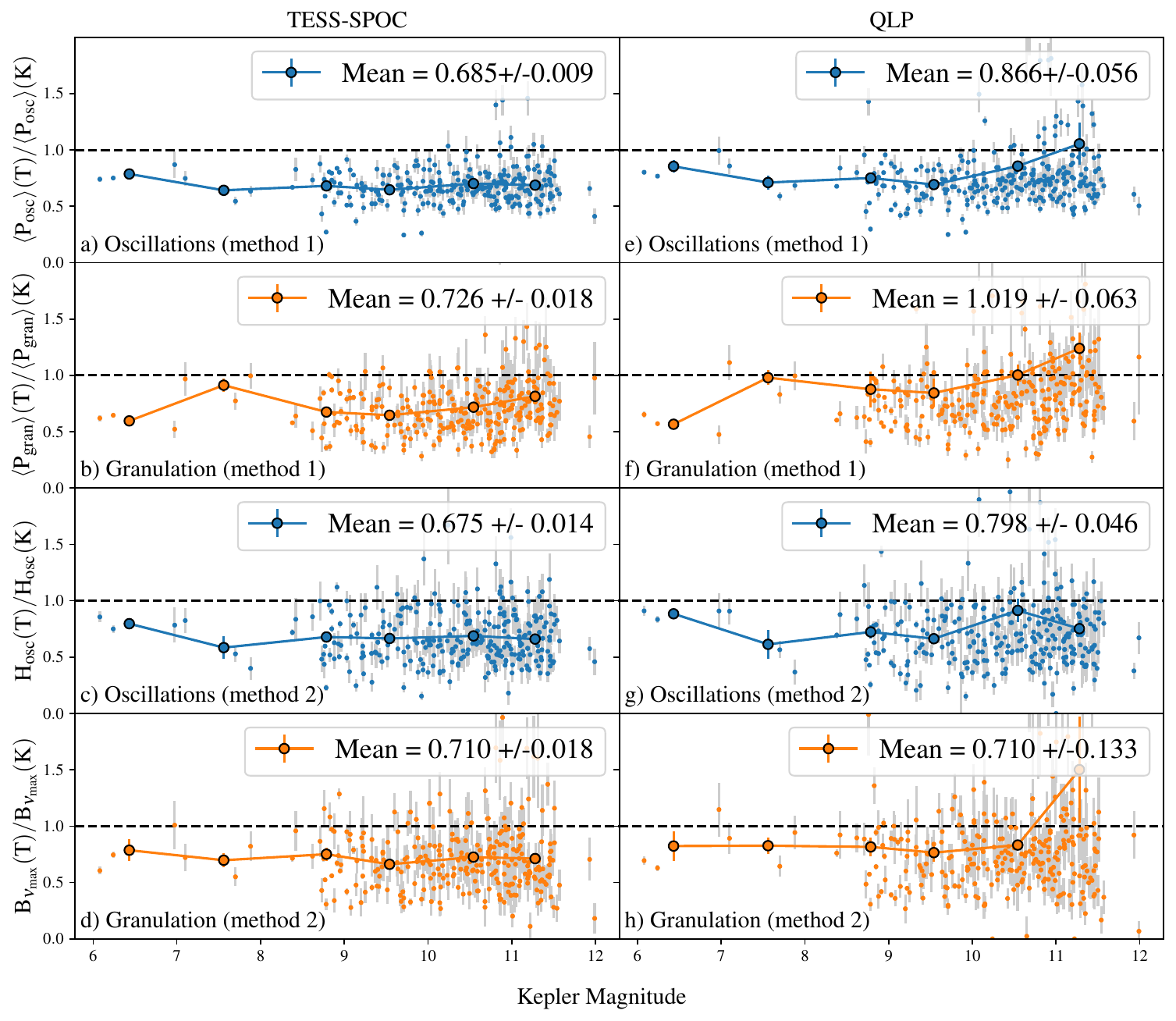}
 \caption{Ratios of oscillation (panel a and c) and ratios of granulation power in TESS to \kepler{} (panel b and d), between TESS and \kepler{}. Plotted as a function of \kepler{} magnitude for red giants. \new{Error bars are shown in grey.} a)~ratio of mean oscillation power \mosc{}. b) ratio of mean granulation power \mgran{}. c)~ratio of oscillation power at \numax{}, \hosc{}. d)~ratio of granulation power at \numax{}\,\bnumax{}. \new{Panels e) - h) shows the same derived for QLP light curves. Note that 12, 26, 7 and 18 stars fall off the top of the plot in panels (e), (f), (g) and (h), respectively. } }
 \label{fig:ratio}
\end{figure*}

\begin{figure}
 \includegraphics[width = \linewidth]{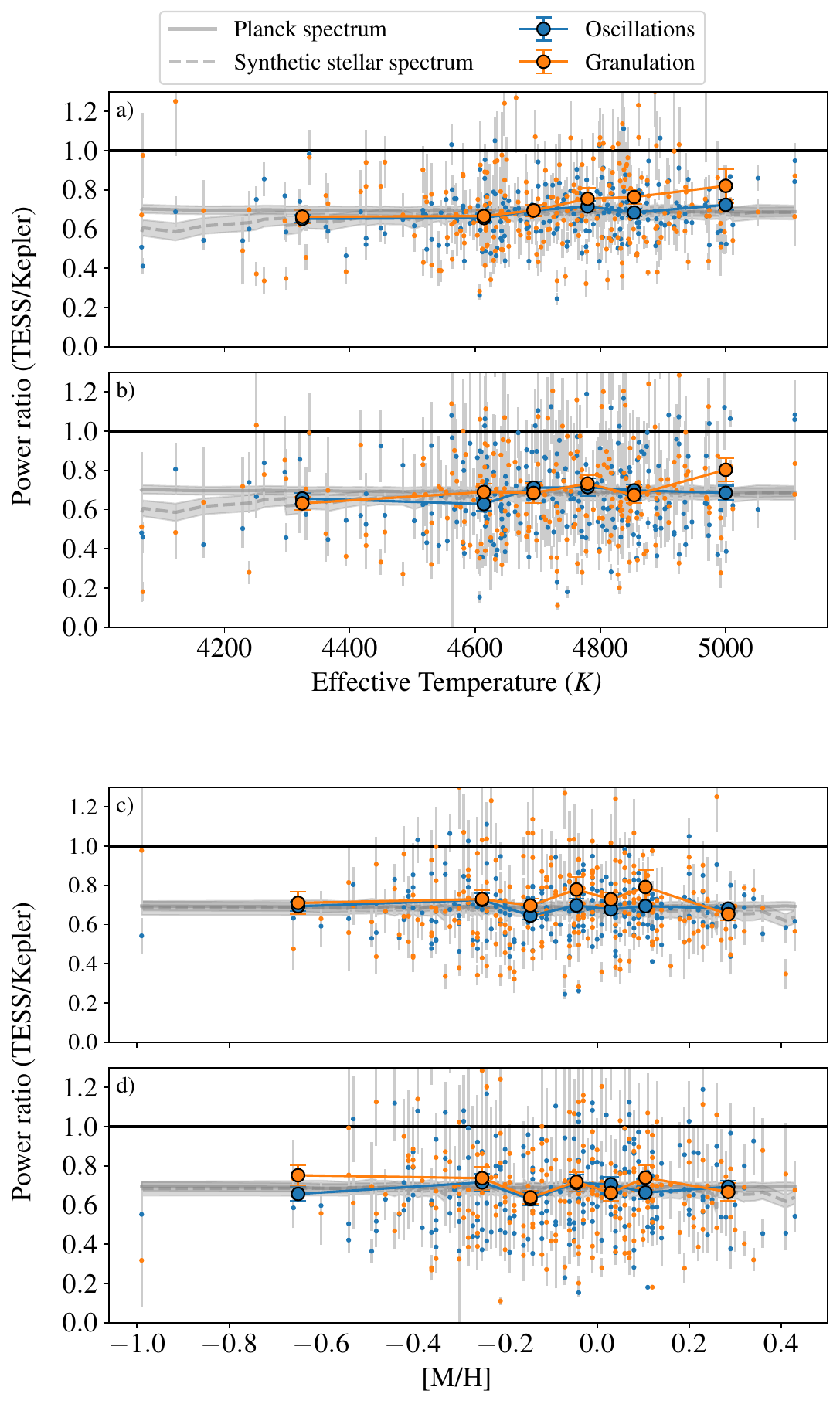}
 \caption{Variation in ratios of oscillation and granulation between TESS and \kepler{}, as a function of stellar parameters. Blue circles represent the ratio of oscillation power in TESS to \kepler{}, orange represent the same for granulation \new{and error bars are shown in grey}. Larger circles show the mean in each bin, and error bars show the SEM. The solid and dashed grey lines represent the predictions from \citet{lund2019} based on Planck spectrum and stellar synthetic spectra, respectively. The grey region around these lines show the error bars for the predicted ratios. a)~ratios of mean power (method 1) as a function of effective temperature. b)~ ratios of fitted parameters (method 2) as a function of effective temperature. c)~ratios of mean power (method 1) as a function of metallicity. d)~ratios of fitted parameters (method 2) as a function of metallicity. }
 \label{fig:teff}
\end{figure}

\subsection{Dependence of oscillation and granulation powers on  \texorpdfstring{$\nu_{\rm max}$}{numax}}

Figure \ref{fig:powers} shows the oscillation power (\mosc{} and \hosc{}) and granulation powers (\mgran{} and \bnumax{}) as a function of \numax{}, from \kepler{} (in purple) and TESS (yellow).
We fitted a polynomial of the form $a\numax{}^{b}$ to these powers, and the results are shown in Table \ref{tab:tab1}. The \mosc{} value (method 1) does not represent an exact measurement of oscillation power, since it also includes a contribution from the background granulation power. In comparison, the \hosc{} from fitting Eq. \ref{eq:eq1} to region 2 is a more accurate measure of oscillation power. The value of $b$ for the \hosc{} fits ($-2.42$ for \kepler{} and $-2.37$ for TESS) are close to the exponents listed in \cite{mosser2012}, which were based only on the first 590 days of \kepler{} data. 
Fig. \ref{fig:powers}b shows the correlation of \mgran{} with \numax{}. We find that \mgran{} $\propto$ \numax{}$^{-2.37}$ for \kepler{} and \mgran{} $\propto$ \numax{}$^{-2.28}$ for TESS, confirming the relation between \numax{} and granulation power from previous studies \citep{mathur2011, mosser2012}. The \bnumax{} values also show this correlation with \numax{},\,\bnumax{} $\propto$ \numax{}$^{-2.19}$. \new{The oscillation powers from the two methods (\mosc{} and \hosc{}) show strong correlation, for both Kepler  and TESS, and the same is true for the granulation power from the two methods (\mgran{} and \bnumax{}). This indicates that both methods provide reasonably good measurements of oscillation and granulation.}

\new{The error bars (in grey) were derived using the Monte Carlo method described by \citet{Huber_2011} \citep[see also][]{mathur2011, regulo2016, sreenivas2024}. Here, we added noise from a $\chi^{2}$ distribution with two degrees of freedom to the power density spectra and repeated methods 1 and 2 for 100 times. The standard deviation of each parameter from this is reported as the error bar. The measurements from \kepler{} generally have a typical median relative uncertainty of 2 to 4\,\%, and 9 to 20\,\% for TESS measurements.}



\subsection{Ratios between TESS and \kepler{}}
\subsubsection{Dependence on observational parameters}
Figure \ref{fig:ratio} shows the ratio of TESS to \kepler{} for oscillation and granulation power, as a function of \kepler{} magnitude. \new{The error bars on individual data points were obtained through propagation of error bars from the previous section. When reporting the final mean, the means from each of the bin were averaged with equal weights.} There are no significant trends in these ratios as a function of brightness, giving confidence that the white noise has been correctly subtracted.
The scatter in the ratios is similar at all brightness levels, suggesting that this scatter does not arise from photon noise and instead arises from the stochastic nature of the processes that excite and damp the oscillations and granulation. This scatter comes predominantly from TESS measurements (see also Fig\,\ref{fig:powers}), although not from the photon noise due to its smaller aperture, but rather because TESS time series are much shorter than those from \kepler{}. Table \ref{tab:results} shows the values of ratios and parameters for 279 stars.
\begin{figure}
 \centering
 \includegraphics[width=0.99\linewidth]{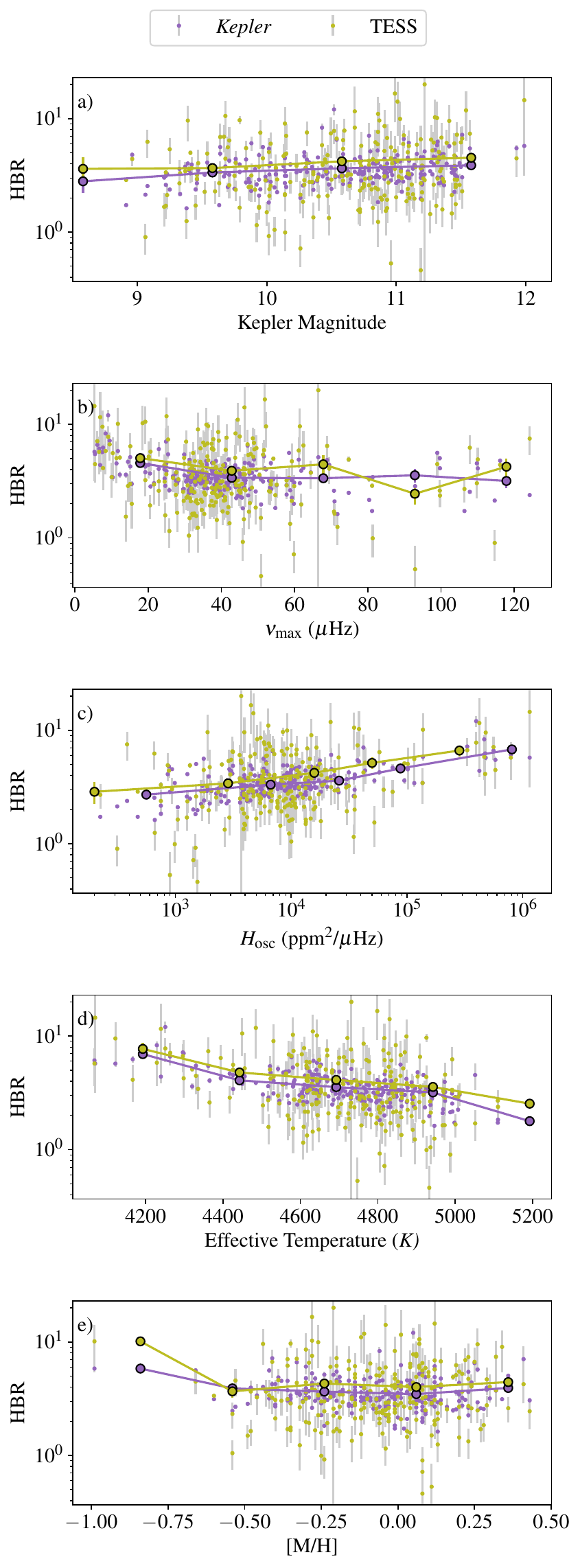}
 \caption{The height-to-background ratio (HBR) from \kepler{} (purple) and TESS (yellow) as a function of various parameters. a)~as a function of \kepler{} magnitude. b)~as a function of \numax{}. c)~as a function of oscillation power at \numax{}, \hosc{}. d)~as a function of effective temperature. e)~as a function of stellar metallicity.}
 \label{fig:hbr}
\end{figure}
 \begin{table*}
\addtolength{\tabcolsep}{-4pt}
    \caption{Table of measurements and parameters used for this work. Only 20 randomly chosen entries are shown here; the full table is available online.}
    \input{figures/table4}
    \label{tab:results}
\end{table*}
 
Fig. \ref{fig:ratio}a shows the ratio of mean oscillation power \mosc{} between \kepler{} and TESS, measured using method 1 (see Sec \ref{subsectmethod1}). We tested for any effect on the results from changing the boundaries of the regions used to measure mean power. We changed these boundaries by 20\,\%, observing that the ratios in each case still agreed with each other within the Standard Error on the Mean.  Fig. \ref{fig:ratio}c shows the ratios of \hosc{} values, which were obtained from fitting a model to the power density spectrum (method 2). The mean ratios from the two methods agree, which is consistent with the fact that the exponents of the power law fit to \mosc{} and \hosc{} are very similar for these two, both in \kepler{} and TESS. \new{The \mosc{} ratios have a typical median relative uncertainty of 9.3\,\%, whereas for \hosc{} it increases to 18\,\%. These large uncertainties are mainly due to the TESS data. } The predicted value of oscillation power ratio between TESS and \kepler{} is $\sim$ 0.69 \citep{lund2019}. We see a mean value of $0.68 \pm 0.01$ for the \mosc{} ratio and $0.68 \pm 0.01$ for the \hosc{} ratio, is excellent agreement with theoretical expectations. 

Similarly, Figs. \ref{fig:ratio}b and d show the ratio of granulation power between TESS and \kepler{}, measured using both methods. The mean ratio of \mgran{} for TESS to \kepler{} is found to be $0.73 \pm 0.02$ (panel b) and the mean ratio of \bnumax{} is $0.71 \pm 0.02$. 
Most importantly, the results confirm that oscillation and granulation have similar wavelength dependence. On comparing \hosc{} (panel c) and \bnumax{} (panel d), it is evident that the granulation power at \numax{} behaves similar to granulation power at low frequencies, at both wavelengths.


For the 7\,\% of stars in our sample that have multiple TESS sectors with gaps, it is interesting to check whether the window function influences these ratios. For this, we also computed the ratios using only the first available sector of each star, and found that they are in good agreement (within 1 SEM) with the values of ratios obtained from the full light curve. 
We also checked how these ratios depend on observational length of the \kepler{} data. For this, we cut out a segment from the \kepler{} time series enforcing it to have the same time span as effective observation time of the used TESS data. We then followed the same methods as before, to calculate the oscillation and granulation powers. The ratios between \kepler{} and TESS have much larger scatter, with means of the ratios unchanged within uncertainties. 

\new{ Figs. \ref{fig:ratio} e to h show the same ratios measured using QLP light curves from the same sectors as TESS-SPOC. We see that the mean values of the ratios from QLP light curves agree with TESS-SPOC for the brightest stars but start to deviate as the stars get fainter and they have an overall large scatter compared to TESS-SPOC light curves. We have found that QLP light curves of some fainter stars contain a set of peaks centered around multiples of one cycle per day (11.57\,$\rm \mu$Hz), due to improper correction of the "Earthshine" effect \citep{conny24}. These peaks have amplitudes similar to oscillations and granulation in our red giants and have influenced the amplitude measurements. Note that, we did not see the effect in QLP measurements of contact binaries (sec. \ref{cbs}) because  the amplitudes of binaries are very large compared to Earthshine artifacts.




}

\subsubsection{Dependence on stellar parameters}

Our measurements allow us to check whether the ratios of oscillation and granulation power between the two pass-bands depend on stellar parameters. The theoretical study by \cite{lund2019} suggested that the oscillation ratio depends on effective temperature and metallicity, with $\log g$ having very little effect. For this analysis, we used the effective temperatures and metallicities from \cite{Yu_2023}, which contains stellar parameters for the 235 stars in our sample.  Figure \ref{fig:teff} shows the different ratios as a function of effective temperature (panels a and b) and metallicity (panels c and d). 
We do not see any dependence of the ratios on these parameters.

We next checked how the measured oscillation ratios agree with the predicted ratios. Using the \new{stellar parameters and corresponding errorbars from \cite{Yu_2023}}, we derived the theoretical ratios using the formalism by \cite{lund2019}, which predicts that the ratios have a slight dependence on effective temperature (\Teff) and stellar metallicity. Over a range of \Teff{} from 4000\,K to 5400\,K, the ratio in power monotonically decreases from 0.698 to 0.686 as per the black body assumption (solid grey line fig. \ref{fig:teff}). On the other hand, the predicted values using synthetic stellar spectra increase from 0.58 to 0.68 until 5000\,K and starts to decrease for hot temperatures beyond this point (dashed grey line fig. \ref{fig:teff}). We note that both predictions agree in the temperature range from 4200\,K to 5800\,K.  \new{Note that there is a sudden jump in ratios from 0.65 to 0.61 at a temperature of about 4296\,K due a higher metallicity star. At a given temperature, such lowering of ratio for higher metallicity stars has been predicted by \cite{lund2019}. } Since the scatter on the observed ratios is larger than this predicted variation, we do not expect this be visible here. It can be seen from Fig.\,\ref{fig:teff} that the measured oscillation ratios follows the theoretical predictions from \cite{lund2019}, for both predictions from using Planck spectrum and using to the stellar model atmospheres, according to \cite{lund2019}. The measured granulation power ratios also agree with these same predictions. 
Finally, we also checked how these ratios depend upon the evolutionary state and we saw no significant difference.

\subsection{Ratio between oscillation and granulation at \numax{}}

Our measurements allow us to check how the ratio between oscillation and granulation power in the region of \numax{} behaves as a function of wavelength. The ratio is essentially a measure of the signal-to-noise of the oscillation signal, and is referred to as the height-to-background ratio (HBR).  Using \kepler{} data spanning 590 days of 1295 giants, \cite{mosser2012} found shown that the HBR, $\frac{\hosc{}}{\bnumax{}}$, is constant at all values of \numax{}, since both \hosc{} and \bnumax{} have the same dependence on \numax{}. However, \cite{kallinger2014} showed that the ratio of granulation amplitude to oscillation amplitudes varies.

For the 235 stars in our sample with stellar parameters, we used \hosc{} and \bnumax{} to compute the HBR. Figure \ref{fig:hbr} shows the variation of HBR over different parameters, for \kepler{} (purple) and TESS (yellow). The HBR  has much less scatter for \kepler{}, again reflecting that the shorter-duration TESS measurements are more affected by the stochastic variations of oscillation and granulation. \new{This can also be seen from the much lower relative uncertainties for \kepler{} data (median 4.6\,\%) compared to 27.4\,\% for TESS data. } Panel a shows that the HBR has no trend with \kepler{} magnitude, both for \kepler{} and TESS. The same is the case for HBR vs \numax{}, except at lower \numax{} (panel b). Panel c shows a clear variation of HBR as a function of oscillation power, \hosc{}. On an average, we see that the HBR has a steady increase towards higher values of \hosc{} for both \kepler{} and TESS, showing the same trend as observed by \cite{kallinger2014}. We see a slight decrease in HBR as we move to higher temperatures for both \kepler{} and TESS (panel d). However, we do not see such an effect when plotted against metallicity, in panel e. In any case, the mean HBR values of TESS does not go below \kepler{} values and suggest asteroseismic signal-to-noise does not depend on wavelength. 

\section{Conclusions}
We have presented the first study of the wavelength dependence of oscillation and granulation powers in a sample of red giants, using light curves from \kepler{} and TESS. Using the publicly available light curves of 51 contact binaries and 279 red giants, we arrive at the following conclusions:
\begin{itemize}
\item Using the amplitudes of contact binary stars, we find that the method used for light curve extraction and de-trending impacts the amplitudes of binary signal. Therefore, careful selection of light curves is important for amplitude studies.
  
 \item The observed value of mean ratio of oscillation power in TESS to \kepler{} agrees with theoretical predictions from \cite{lund2019}. The mean oscillation power \mosc{} and power density at \numax{} (\hosc{}) ratios both agree with prediction, with mean values $0.68\,\pm\,0.01$. A global measure of granulation power (\mgran{}) also shows good agreement with predicted ratios. We see no significant departure to this ratio at any \kepler{} magnitude or time span of TESS-SPOC data. 

 \item Oscillation and granulation have the same wavelength dependence, as predicted by \cite{kjb11}. That is, the granulation signal will reduce in the same way as the oscillation signal does as we observe through redder wavelengths. 

 \item The mean oscillation and granulation ratios from observations do not show any significant dependence on effective temperature or metallicity. This shows that observations and corresponding measurements at redder wavelengths do not show any bias towards \new{red giants}.

 \item The height-to-background ratio (HBR) of oscillation and granulation at \numax{} do not depend on wavelength, at least over the limited baseline.  This is reassuring for the Roman mission, which will observe at even redder wavelengths, giving no reason to worry that the oscillations might be swamped by granulation. 
 
\end{itemize}


\begin{figure}
 \centering
 \includegraphics[width=0.99\linewidth]{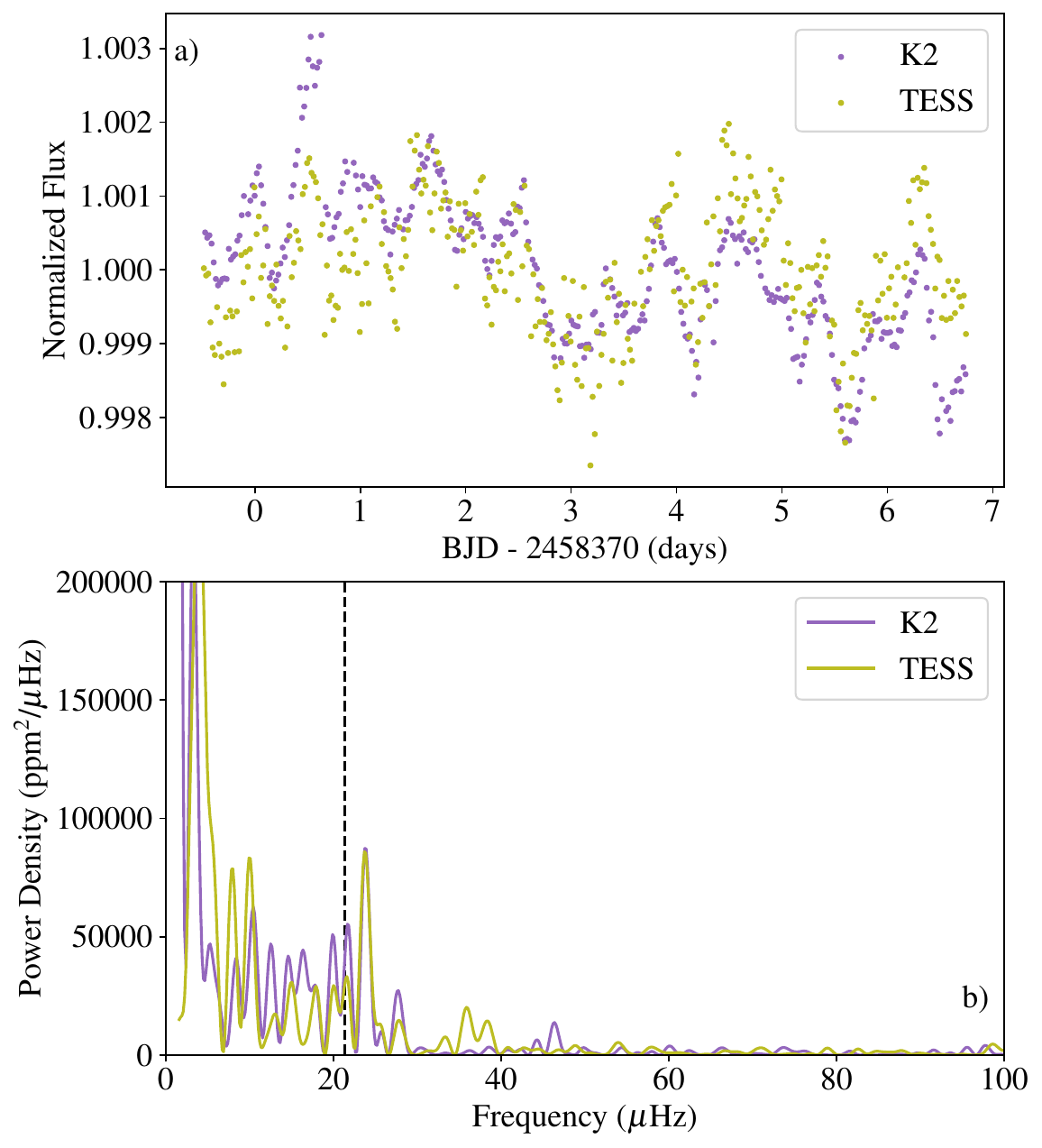}
 \caption{Simultaneous observations of the red giant EPIC\,245926259 (TIC\,404244638) from K2 (purple) and TESS (yellow). a) The light curves b) The power density spectra. The vertical dashed line shows the \numax{} = 21.89\,$\mu$Hz from \citet{k2gap2022}, based on full K2 light curve. }
 \label{fig:k2tess}
\end{figure}
Additional data will be required to put more stringent observational constraints on the wavelength dependence of granulation and oscillations. An excellent way to do this would be using simultaneous observations in two wavelength regions, since the stochastic variations of oscillations and granulation will be the same in both light curves. The BRITE-Constellation satellite mission (BRIght Target Explorer; \citealt{brite2014}) could in principle do this, since it has detected oscillations in some bright red giants \citep{kallinger2019} and it has red and blue filters. Meanwhile, a nice example comes from noting that \cite{thomask2} found 171 stars that were observed simultaneously  for 8 days by K2 (campaign 19) and TESS (sector 2). Most are faint main-sequence stars but their list does contain four red giants with oscillations reported by \cite{k2gap2022} using the full K2 dataset. 
Figure \ref{fig:k2tess} shows the best example, with the K2 light curve from K2SFF \citep{k2sff} and the TESS light curve from QLP \citep{qlp1,qlp2} and the corresponding power density spectra. Even though the data spans only 8 days, the same stellar variability can be clearly seen in both light curves (panel a) and the PDS are also very similar (panel b).  A longer data set would have allowed the power ratio to be measured accurately and this example is promising for the PLATO mission \citep{plato24}, which will observe stars simultaneously using its red and blue fast cameras. Finally, we note that the infrared light curves being obtained for the study of exoplanet atmospheres from JWST \citep[e.g.][]{jwst} may also allow a direct measurement of granulation amplitudes at infrared wavelengths, even before the planned launch of the Roman mission.


%



%

\section*{Acknowledgements}
\new{We thank Mikkel N. Lund, Benjamin Montet and Andrej Pr\v{s}a for useful discussions, and Simon J. Murphy for comments on the draft manuscript.} 
We acknowledge support from the Australian Research Council through Laureate Fellowship FL220100117, which includes a PhD scholarship for KRS. 
D.H. acknowledges support from the Alfred P. Sloan Foundation, the National Aeronautics and Space Administration (80NSSC19K0597, 80NSSC21K0652) and the Australian Research Council (FT200100871). MGP is supported by the Professor Harry Messel Research Fellowship in Physics Endowment, at the University of Sydney. This paper includes data collected by the \kepler{} mission and obtained from the MAST data archive at the Space Telescope Science Institute (STScI). Funding for the Kepler mission is provided by the NASA Science Mission Directorate. STScI is operated by the Association of Universities for Research in Astronomy, Inc., under NASA contract NAS 5–26555. This paper includes data collected with the TESS mission, obtained from the MAST data archive at the Space Telescope Science Institute (STScI). Funding for the TESS mission is provided by the NASA Explorer Program. STScI is operated by the Association of Universities for Research in Astronomy, Inc., under NASA contract NAS 5–26555.
This work made use of several publicly available {\tt python} packages: {\tt astropy} \citep{astropy:2013,astropy:2018}, 
{\tt lightkurve} \citep{lightkurve2018},
{\tt matplotlib} \citep{matplotlib2007}, 
{\tt numpy} \citep{numpy2020}, and 
{\tt scipy} \citep{scipy2020}. 
This work has made an extensive use of
Topcat (\url{https://www.star.bristol.ac.uk/~mbt/topcat/}).

\section*{Data Availability}

The \kepler{} and TESS data underlying this article are available at the MAST Portal (Barbara A. Mikulski Archive for Space Telescopes), at \url{https://mast.stsci.edu/portal/Mashup/Clients/Mast/Portal.html}

\ifarxiv
 \input{output.bbl} 
\else
 \bibliographystyle{mnras}
 \bibliography{references}
\fi

\appendix

\bsp	
\label{lastpage}
\end{document}

%% file: figures/table4.tex
\begin{tabular}{cccccccccccc}
\toprule
KIC & TIC & K$_{p}$ & \numax{} & Number of & \mosc{}  & \mgran{} & \hosc{} & \bnumax{}  & ratio  & ratio \\
    &             &    &     ($\rm\mu$Hz) & TESS-SPOC sectors   &  ratio    &    ratio      &  ratio     &      ratio      &  planck spectrum & stellar spectra \\ 
\midrule
11042388 & 28087079 & 11.51 & 9.492 & 1 & 0.769\,$\pm$\,0.010 & 0.64\,$\pm$\,0.23 & 0.79\,$\pm$\,0.16 & 0.85\,$\pm$\,0.18 & 0.697\,$\pm$\,0.044 & 0.655\,$\pm$\,0.069 \\
12110800 & 26473705 & 11.51 & 38.93 & 1 & 0.90\,$\pm$\,0.10 & 0.78\,$\pm$\,0.11 & 0.87\,$\pm$\,0.18 & 1.16\,$\pm$\,0.24 & 0.691\,$\pm$\,0.043 & 0.680\,$\pm$\,0.068 \\
11621486 & 351191961 & 11.5 & 37.54 & 1 & 0.478\,$\pm$\,0.042 & 0.93\,$\pm$\,0.15 & 0.461\,$\pm$\,0.076 & 0.280\,$\pm$\,0.092 & 0.691\,$\pm$\,0.046 & 0.686\,$\pm$\,0.074 \\
11147596 & 27773596 & 11.49 & 22.89 & 1 & 0.588\,$\pm$\,0.0597 & 0.730\,$\pm$\,0.147 & 0.448\,$\pm$\,0.079 & 0.70\,$\pm$\,0.14 & 0.696\,$\pm$\,0.042 & 0.632\,$\pm$\,0.081 \\
11090395 & 27239115 & 11.47 & 39.33 & 1 & 0.686\,$\pm$\,0.064 & 0.559\,$\pm$\,0.086 & 0.88\,$\pm$\,0.12 & 0.372\,$\pm$\,0.096 & 0.688\,$\pm$\,0.046 & 0.686\,$\pm$\,0.078 \\
11152312 & 28362724 & 11.47 & 35.49 & 1 & 0.791\,$\pm$\,0.066 & 0.65\,$\pm$\,0.11 & 0.73\,$\pm$\,0.12 & 0.78\,$\pm$\,0.15 & 0.690\,$\pm$\,0.045 & 0.680\,$\pm$\,0.071 \\
11358755 & 27642828 & 11.46 & 37.48 & 1 & 0.591\,$\pm$\,0.052 & 0.560\,$\pm$\,0.092 & 0.421\,$\pm$\,0.073 & 0.77\,$\pm$\,0.16 & 0.692\,$\pm$\,0.044 & 0.667\,$\pm$\,0.070 \\
12302104 & 298962109 & 11.46 & 34.68 & 1 & 0.614\,$\pm$\,0.063 & 0.98\,$\pm$\,0.15 & 0.47\,$\pm$\,0.11 & 0.84\,$\pm$\,0.13 & 0.688\,$\pm$\,0.046 & 0.686\,$\pm$\,0.075 \\
10416550 & 272279790 & 11.45 & 49.03 & 1 & 0.834 \,$\pm$\,0.068 & 0.928\,$\pm$\,0.135 & 0.95\,$\pm$\,0.11 & 0.55\,$\pm$\,0.11 & 0.69\,$\pm$\,0.044 & 0.684\,$\pm$\,0.071 \\
10816214 & 28450479 & 11.45 & 47.65 & 1 & 0.737\,$\pm$\,0.071 & 0.93\,$\pm$\,0.14 & 0.68\,$\pm$\,0.15 & 0.65\,$\pm$\,0.20 & 0.689\,$\pm$\,0.046 & 0.689\,$\pm$\,0.070 \\
11567572 & 27648781 & 11.45 & 59.25 & 1 & 0.850\,$\pm$\,0.084 & 0.486\,$\pm$\,0.070 & 1.19\,$\pm$\,0.23 & 0.57\,$\pm$\,0.22 & 0.690\,$\pm$\,0.046 & 0.675\,$\pm$\,0.070 \\
10492041 & 268384311 & 11.43 & 46.15 & 1 & 0.808\,$\pm$\,0.072 & 1.02\,$\pm$\,0.16 & 0.55\,$\pm$\,0.13 & 1.37\,$\pm$\,0.19 & 0.692\,$\pm$\,0.045 & 0.688\,$\pm$\,0.073 \\
11357931 & 27457606 & 11.43 & 37.36 & 1 & 0.467\,$\pm$\, 0.039 & 0.466\,$\pm$\,0.064 & 0.353\,$\pm$\,0.074 & 0.56\,$\pm$\,0.10 & 0.689\,$\pm$\,0.046 & 0.690\,$\pm$\,0.074 \\
10292064 & 273682689 & 11.41 & 7.521 & 1 & 0.69\,$\pm$\,0.10 & 1.25\,$\pm$\,0.44 & 0.80\,$\pm$\,0.17 & 0.48\,$\pm$\,0.13 & 0.700\,$\pm$\,0.042 & 0.586\,$\pm$\,0.089 \\
9852480 & 268614931 & 11.39 & 37.38 & 1 & 0.734\,$\pm$\,0.069 & 1.137\,$\pm$\,0.199 & 0.769\,$\pm$\,0.119 & 0.50\,$\pm$\,0.15 & 0.688 \,$\pm$\,0.043 & 0.686\,$\pm$\,0.072 \\
11414700 & 28085491 & 11.37 & 39.63 & 1 & 0.599\,$\pm$\,0.069 & 0.84 \,$\pm$\,0.13 & 0.67\,$\pm$\,0.13 & 0.62\,$\pm$\,0.12 & 0.690\,$\pm$\, 0.045 & 0.683\,$\pm$\,0.072 \\
10224535 & 273377694 & 11.36 & 38.81 & 1 & 0.592\,$\pm$\,0.055 & 0.706\,$\pm$\,0.099 & 0.56\,$\pm$\,0.11 & 0.52\,$\pm$\,0.12 & 0.690\,$\pm$\,0.046 & 0.679\,$\pm$\,0.071 \\
10552614 & 273681418 & 11.35 & 46.43 & 1 & 0.515\,$\pm$\,0.051 & 0.60\,$\pm$\,0.10 & 0.50\,$\pm$\,0.084 & 0.418\,$\pm$\,0.096 & 0.688\,$\pm$\,0.045 & 0.688\,$\pm$\,0.071 \\
11508970 & 26656546 & 11.35 & 36.21 & 1 & 0.658\,$\pm$\,0.076 & 0.74\,$\pm$\,0.11 & 0.85\,$\pm$\,0.14 & 0.34\,$\pm$\,0.10 & 0.69\,$\pm$\,0.044 & 0.695\,$\pm$\,0.071 \\
12264089 & 27231412 & 11.34 & 41.97 & 1 & 0.466\,$\pm$\,0.046 & 0.579\,$\pm$\,0.077 & 0.358\,$\pm$\,0.070 & 0.51\,$\pm$\,0.10 & 0.693\,$\pm$\, 0.044 & 0.687\,$\pm$\,0.072 \\
\bottomrule
\end{tabular}